\documentclass[twocolumn]{aastex631}

\usepackage{graphicx}	
\usepackage{amsmath}	
\usepackage{booktabs}  
\usepackage{natbib}
\usepackage{float}
\usepackage{hyperref}

\begin{document}
\title{Challenging Classical Paradigms: Recurrent Nova M31N 2017-01e, a BeWD system in M31?}
\correspondingauthor{Shatakshi Chamoli}
\email{shatakshi.chamoli@iiap.res.in, shatakshichamoli@gmail.com}

\author[0009-0000-5909-293X]{Shatakshi Chamoli}
\affiliation{Indian Institute of Astrophysics, 2nd Block Koramangala, 560034, Bangalore, India}
\affiliation{Pondicherry University, R.V. Nagar, Kalapet, 605014 Puducherry, India}

\author[0000-0001-7570-545X]{Judhajeet Basu}
\affiliation{Indian Institute of Astrophysics, 2nd Block Koramangala, 560034, Bangalore, India}
\affiliation{Pondicherry University, R.V. Nagar, Kalapet, 605014 Puducherry, India}

\author[0000-0002-3927-5402]{Sudhanshu Barway}
\affiliation{Indian Institute of Astrophysics, 2nd Block Koramangala, 560034, Bangalore, India}

\author[0000-0003-3533-7183]{G.C. Anupama}
\affiliation{Indian Institute of Astrophysics, 2nd Block Koramangala, 560034, Bangalore, India}

\author[0000-0002-7942-8477]{Vishwajeet Swain}
\affiliation{Department of Physics, Indian Institute of Technology Bombay, Powai, Mumbai 400076, India}

\author[0000-0002-6112-7609]{Varun Bhalero}
\affiliation{Department of Physics, Indian Institute of Technology Bombay, Powai, Mumbai 400076, India}

\begin{abstract}
M31N 2017-01e is the second-fastest recurrent nova known, with a recurrence period of 2.5 years in the Andromeda Galaxy (M31). This system exhibits a unique combination of properties: a low outburst amplitude ($\sim3$ magnitude), starkly contrasting with known recurrent novae (typically $\geq 6$ magnitudes), and a very fast evolution ($t_{2}\sim 5 $ days). Its position coincides with a bright variable source ($\mathrm{{M_V \sim -4.2,\, B-V= 0.042}}$) displaying a 14.3 day photometric modulation, which has been suggested as the likely progenitor. 
We present a multi-wavelength analysis of optical and UV data spanning quiescence and the 2019 and 2024 outbursts. Archival high-resolution imaging reveals two nearby faint sources within $5\arcsec$ of the proposed nova system, which we identified as unrelated field stars. Color analysis and spectral energy distribution fitting suggest the progenitor is likely an early-type star. Combined with archival spectra consistent with a B-type star with H$\alpha$ in emission, this points to the quiescent counterpart being a Be star with a circumstellar disc.  
We propose that M31N 2017-01e arises from a {\it rare Be-WD binary}, where the WD accretes from the decretion disk of its companion, explaining its rapid recurrence, low-amplitude outbursts, and unusual quiescent luminosity and color. This analysis highlights M31N 2017-01e as a compelling outlier among recurrent novae, suggesting a distinct accretion mechanism and evolutionary path that challenges the prevailing paradigm.
\end{abstract}


\section{Introduction}
 \begin{table*}[t!]
    \centering
    \caption{Eruption history of M31N 2017-01e.}
    \begin{tabular}{ccccc}
    \hline
    \noalign{\vskip 1.2ex} 
    \textbf{Discovery date (UT)} & \textbf{\shortstack{Time since previous \\ eruption (days)}} & \textbf{Name} & \textbf{Observer} & \textbf{References} \\
    \hline
    2012 Jan 11.20 &    --  & PS1\_DR2\_11.04464+41.9061 & Pan-STARRS & [1] \\
    2014 Jun 19.47 & 891.36 & M31N 2014-06c & Pan-STARRS & [2] \\
    2017 Jan 31.40 & 956.93 & TCP J00441072+4154221 & K. Itagaki & [3] \\
    2019 Sep 22.60 & 964.25 & PNV J00441073+4154220 & XOSS & [4]\\
    2022 Mar 05.60 & 894.95 & PNV J00441072+4154224 & XOSS & [5] \\
    2024 Aug 06.80 & 885.24 & PNV J00441071+4154221 & XOSS & [6],[7] \\

    \hline
    \end{tabular}
    References: [1] \cite{shafter-2022}, [2] \cite{shafter-2024}, [3] \cite{K.Itagaki-2017}, [4] \cite{Xoss-2019}, [5] \cite{Xoss-2022}, [6] \cite{shafter-2024}, [7] \cite{git-2024} 
    \label{erp-hist}
\end{table*}

Novae are a class of cataclysmic variables consisting of a white dwarf (WD) and a companion star, which is typically a late-type main-sequence, sub-giant, or a giant star (see \citealt{2021ARA&A..59..391C} for a recent review). The WD accretes matter from the companion either via Roche Lobe Overflow (RLOF) or stellar winds. As matter accumulates on the surface of the WD, a thermonuclear runaway (TNR) leads to an outburst observed as an increase in brightness of the system ($\geq 6$ mags; \citealt{2014ApJ...788..164P}). The accreted envelope is ejected at high velocities ($10^2$-$10^3$ km s$^{-1}$), cooling down as it expands and the system gradually fades. The time taken to decline by two magnitudes from the peak, $t_2$, defines the nova's speed class ranging from very fast ($t_2 < 10$ days) to slow ($t_2 > 100$ days; \citealt{speed-class}). The residual accreted material on the WD burns steadily, with emission peaking in the soft X-rays and extending to the ultraviolet (UV) wavelengths in the Super Soft Source (SSS) phase. Soon after an outburst, the accretion process resumes, and the system can erupt again. For most novae the time taken to show another outburst, the recurrence period, is longer than our observation baseline, these systems are called Classical Novae (CN). Whereas a few systems, known as Recurrent Novae (RN), have multiple observed eruptions over recurrence timescales of decades or less. A few Symbiotic novae, a long period system in which the WD accretes from a cool giant companion, also show recurrent outbursts (\citealt{2025CoSka..55c..47M} and references therein).

M31 hosts novae with the shortest known recurrence period till date. These RNe are grouped into a subclass called rapid recurrent novae (RRN), and are candidates for progenitors of type Ia supernovae via the single degenerate channel \citep{2020AdSpR..66.1147D}. The nova M31N 2008-12a  goes into outburst every year and has been studied in great detail (\citealt{basu_21a,Henze_12a,2016ApJ...833..149D} and references therein). The nova M31N 2017-01e, first discovered by \cite{K.Itagaki-2017}, has the second shortest known recurrence period, erupting approximately every 2.5 years. Spectroscopic observations of its 2017 outburst \citep{atel-spec} revealed broad Balmer lines, along with He I and N III emission lines, characteristic of fast novae and similar to other well-known recurrent novae such as U Sco \citep{2015ApJ...811...32P,2022MNRAS.516.4497S} and LMCN 1968-12a (\citealt{1991ApJ...370..193S,2020MNRAS.491..655K}, Basu et. al under prep). Since its discovery, M31N 2017-01e has been observed in multiple outbursts, and archival searches have identified past eruptions (see Table-\ref{erp-hist}).

Post the 2022 outburst, \citet{shafter-2022} noted that the nova's position is coincident with a bright, blue variable source, with V-band magnitude of $\sim 20.4$ \citep{Massey} exhibiting a periodic modulation of 14.3~days~\citep{Vilardell}. More recently, \citet{shafter-2024} reported the spectra of the quiescent counterpart identifying it as an evolved B-type star with prominent H$\alpha$ emission and He I absorption feature. Since novae have late-type companions, the association with a B-type star is unusual.

In this work, we present a comprehensive photometric analysis of M31N 2017-01e during both quiescence and eruption, with emphasis on the 2019 and 2024 events. We analyze the system's light curve (LC), colors, and spectral energy distribution (SED) to constrain its nature, in particular, the identity of its companion star.

\section{Multi-Wavelength Observations of M31N 2017-01e} \label{obs}
We utilized multi-wavelength observations from ground- and space-based facilities to monitor and characterize the 2024 outburst of the nova. We supplemented our analysis with archival data to examine the 2019 outburst and the quiescent behavior of the system. Photometric measurements were obtained in the optical, UV, and X-ray bands. Data reduction and calibration were performed using standardized pipelines and reference catalogs. In the following, we briefly describe the observing facilities used for this work. 

UV observations of the 2024 outburst were obtained with the \textit{AstroSat/UltraViolet Imaging Telescope} (UVIT; FOV $28\arcmin$, resolution $1.5\arcsec$; \citealt{uvit}) in the F148W filter under ToO proposal T05\_225. We used archival UVIT data in the FUV (F148W, F172M) and NUV (N219M) filters from M31 field 10 \citep{m31-fields} and past observations of M31N 2008-12a nova, where the object is located $\sim 14.5\arcmin$ from the center (see Table-\ref{uvit-table}). The archival \texttt{Level 2} data was downloaded from the Indian Space Science Data Center's (ISSDC) \textit{astrobrowse} portal\footnote{\href{https://astrobrowse.issdc.gov.in/astro_archive/archive/Home.jsp}{https://astrobrowse.issdc.gov.in/astro\_archive/archive/Home.jsp}}. Data reduction followed \citet{ccdlab} in CCDLAB, and PSF photometry was performed using \texttt{daophot} package in IRAF \citep{1986SPIE..627..733T}. Photometric zero-points from \citet{UVIT-cal} were applied along with aperture corrections to account for the extended wings of the UVIT PSF. Special care is taken to mitigate flux contamination from nearby sources. 

The \textit{GROWTH-India Telescope} (GIT; \citealt{GIT-obs}), a 0.7~m facility at the Indian Astronomical Observatory (IAO), Hanle, observed the nova using the SBIG camera (FOV $7\arcmin \times 10\arcmin$) in the SDSS $g^\prime$, $r^\prime$, and $i^\prime$ filters. This nova field was being continuously monitored as part of the GIT M31 survey and followed-up the 2024 outburst of the nova closely. Data processing was done with the automated GIT pipeline, and PSF photometry was calibrated against PanSTARRS DR1 \citep{2016arXiv161205560C}. 

The \textit{Himalayan Chandra Telescope} (HCT) \footnote{\href{https://www.iiap.res.in/centers/iao/facilities/hct/}{https://www.iiap.res.in/centers/iao/facilities/hct/}}, a 2~m telescope at IAO, observed this object with the Himalayan Faint Object Spectrograph (HFOSC) in SDSS-$u^\prime$ and $g^\prime$ bands. Data reduction and PSF photometry were performed in IRAF, with calibration against SDSS-DR12 \citep{2015ApJS..219...12A} (see Table-\ref{2024-opt}).

The \textit{Canada-France-Hawaii Telescope} (CFHT) MegaPrime/MegaCam  \citep{2003SPIE.4841...72B}(FOV $1^\circ \times 1^\circ$, median seeing $0.7\arcsec$) provided optical data in u.MP9301, g.MP9401, r.MP9601 and i.MP9701 filters (hereafter referred to as CFHT-$u$, $g$, $r$ and $i$) from the Canadian Astronomy Data Centre \footnote{\href{https://www.cadc-ccda.hia-iha.nrc-cnrc.gc.ca/en/}{https://www.cadc-ccda.hia-iha.nrc-cnrc.gc.ca/en/}}. We downloaded archival level 2 data within a $1\arcmin$ region around the nova along with level 3 median stacked data generated via the MegaPipe pipeline. In this work we have used data with exposure times greater than 100s, the majority of which were observed over multiple nights in August-October of 2004 and a few nights from 2005, 2009, 2010 and 2012. PSF photometry utilized \texttt{sextractor} \citep{1996A&AS..117..393B} and \texttt{psfex} \citep{2011ASPC..442..435B} for model generation and flux extraction, with calibration following standard procedures\footnote{\href{https://www.cfht.hawaii.edu/Science/CFHTLS-DATA/megaprimecalibration.html}{https://www.cfht.hawaii.edu/Science/CFHTLS-DATA/megaprimecalibration.html}}. A partial magnitude table in presented in Table-\ref{cfht-table} and the full table is available in the electronic format.

\textit{Swift} \citep{Gehrels_2004} observations of the 2019 outburst included the UltraViolet/Optical Telescope (UVOT; \citealt{swift}) in UVW2 and U bands and the X-Ray Telescope (XRT; \citealt{Burrows_2005}). \texttt{Level 2} UVOT data were analyzed using \texttt{uvotsource} in \texttt{HEASoft}, with aperture photometry on a $5\arcsec$ source region and a $35\arcsec$ background region. XRT data, processed with \texttt{sosta}, revealed no significant detection above a $2.8\sigma$ level. The archival data used and the source magnitudes are summarized in Table-\ref{uvot-table}. 

We queried VizieR for cataloged counterparts within $1\arcsec$ of the nova. The source was identified as $M31V\,J00441070+4154220$ in the eclipsing binaries catalog by \citet{Vilardell}, with a 14.3 day periodicity. It is listed in Guide Star Catalog 2.4.2 \citep{gsc_2.4.2,gsc_ref} under ID $NBW9051234$, from which we obtained SDSS-$g^{\prime}$, $r^{\prime}$, $i^{\prime}$, $z^{\prime}$ band photometry. We also included a $y$-band detection from archival Pan-STARRS data \citep{2020ApJS..251....7F} under object-ID $158280110446247958$, at an epoch corresponding to $\sim550$ days post the 2012 eruption \citep{shafter-2024}. We additionally retrieved photometric data for the source $J004410.7+415422$ from the XMM-Newton Serendipitous Ultraviolet Source Survey catalog \citep{2012MNRAS.426..903P}, covering the UVW2, UVM2, UVW1, and U bands. The nova was observed during both the 2019 and 2024 outbursts, as well as during quiescent periods, by the Zwicky Transient Facility (ZTF). LCs in the $g$, $r$, and $i$ bands were obtained from SNAD \citep{snad} and ALeRCE \citep{alerce} platforms.

\begin{figure*}[t!]
    \centering
    \includegraphics[width=\textwidth]{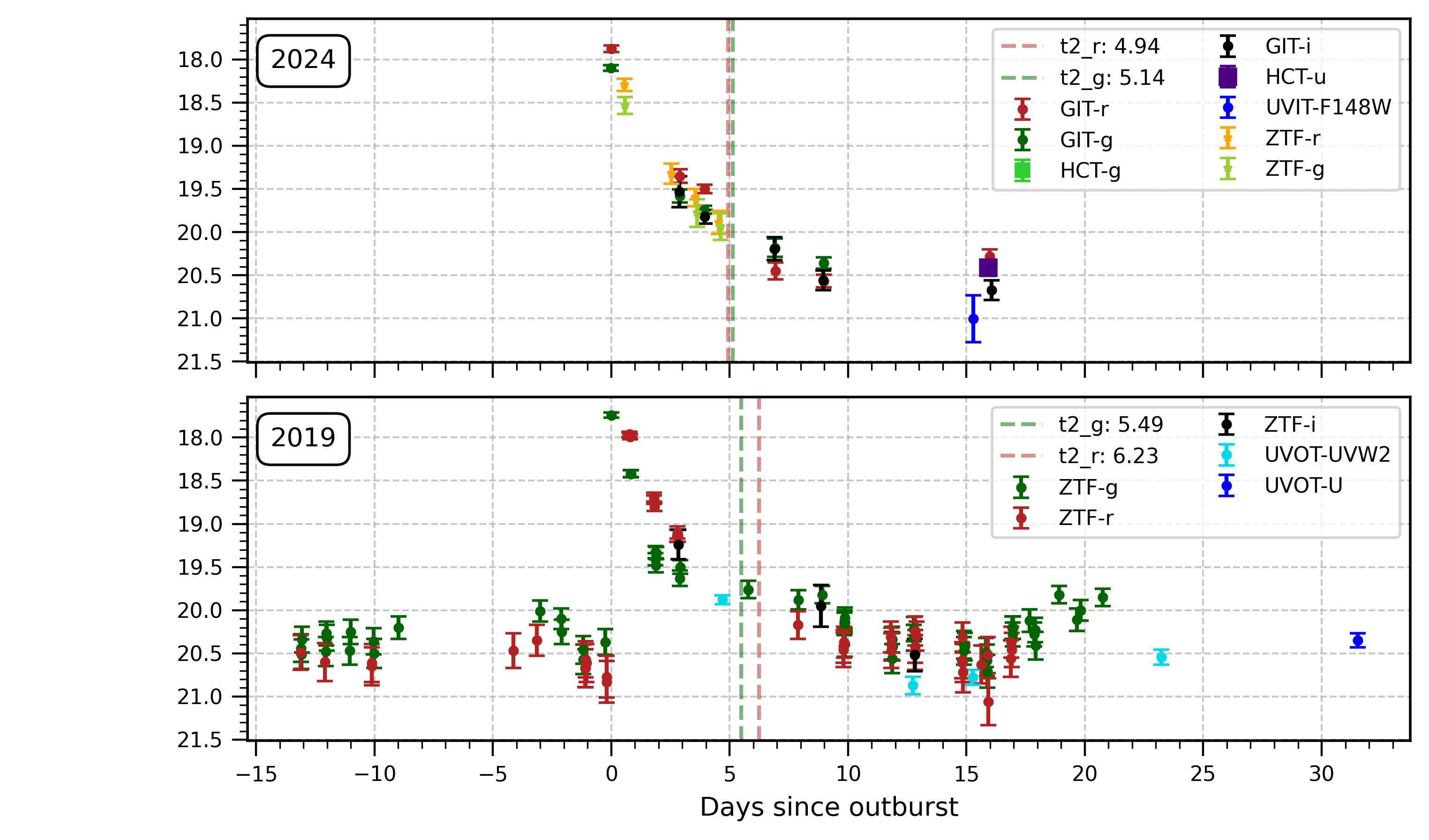}
    \caption{(Top): 2024 outburst light curve, (Bottom): 2019 outburst light curve; $\mathrm{t_{2}}$ is marked in vertical dashed line. The 2024 data is given in Table-\ref{2024-opt}. All ZTF data used is available on SNAD and ALeRCE. The Swift/UVOT data is tabulated in Table-\ref{uvot-table}.}
    \label{fig:lc}
\end{figure*}

\begin{figure*}[t]
    \centering
    \includegraphics[width=1.0\textwidth]{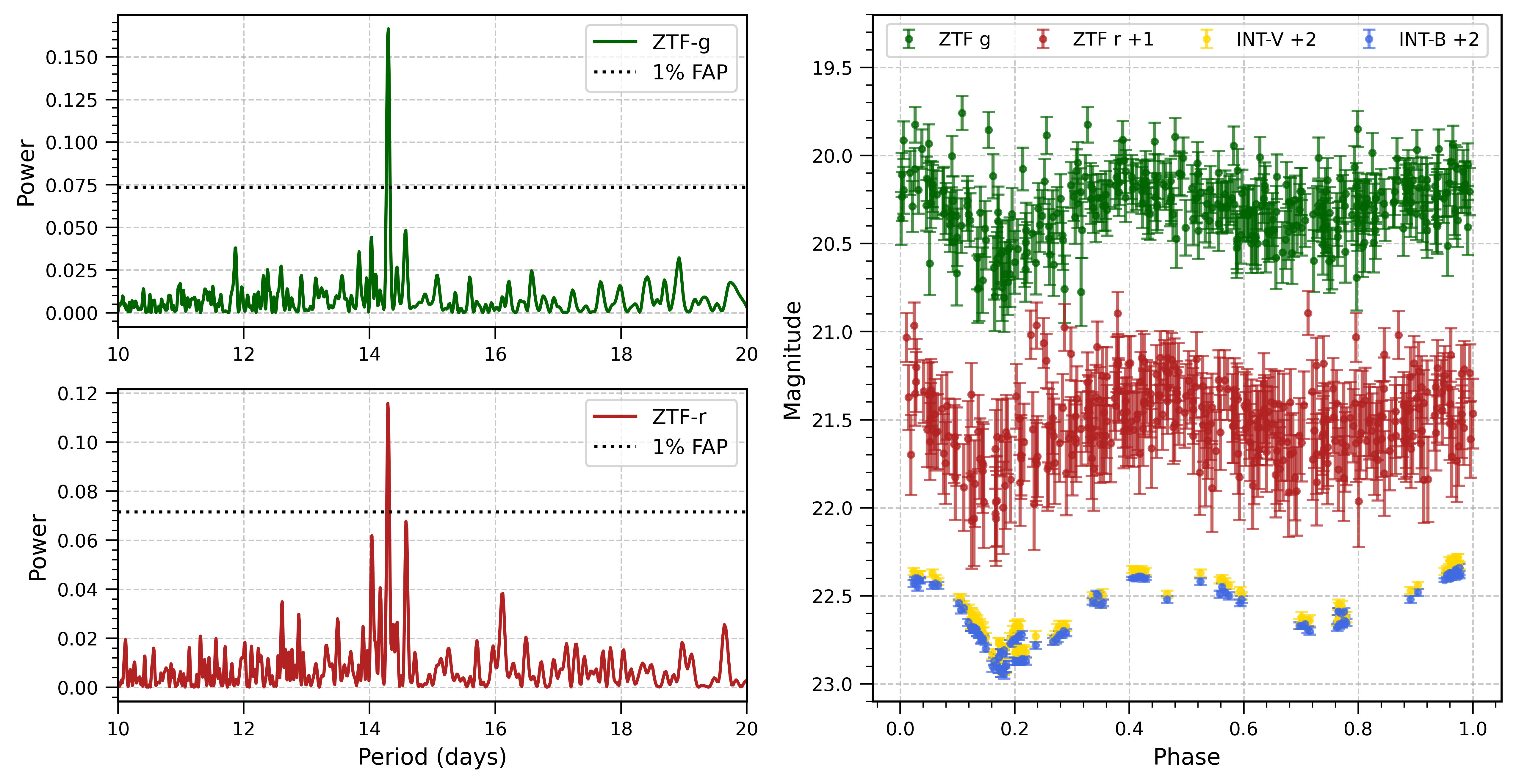}
    \caption{Left: Lomb-Scargle periodogram of the ZTF data, showing the highest power at a period of 14.29 days. The dotted line marks the $1\%$ FAP (False alarm probability) level, indicating the statistical significance of the periodogram peak, Right: Phase-folded Light curve. The ZTF-phase has been shifted by a constant value to match the phase of INT data.}
    \label{lsp}
\end{figure*}

\begin{figure*}[t]
    \centering
    \includegraphics[width=1.0\textwidth]{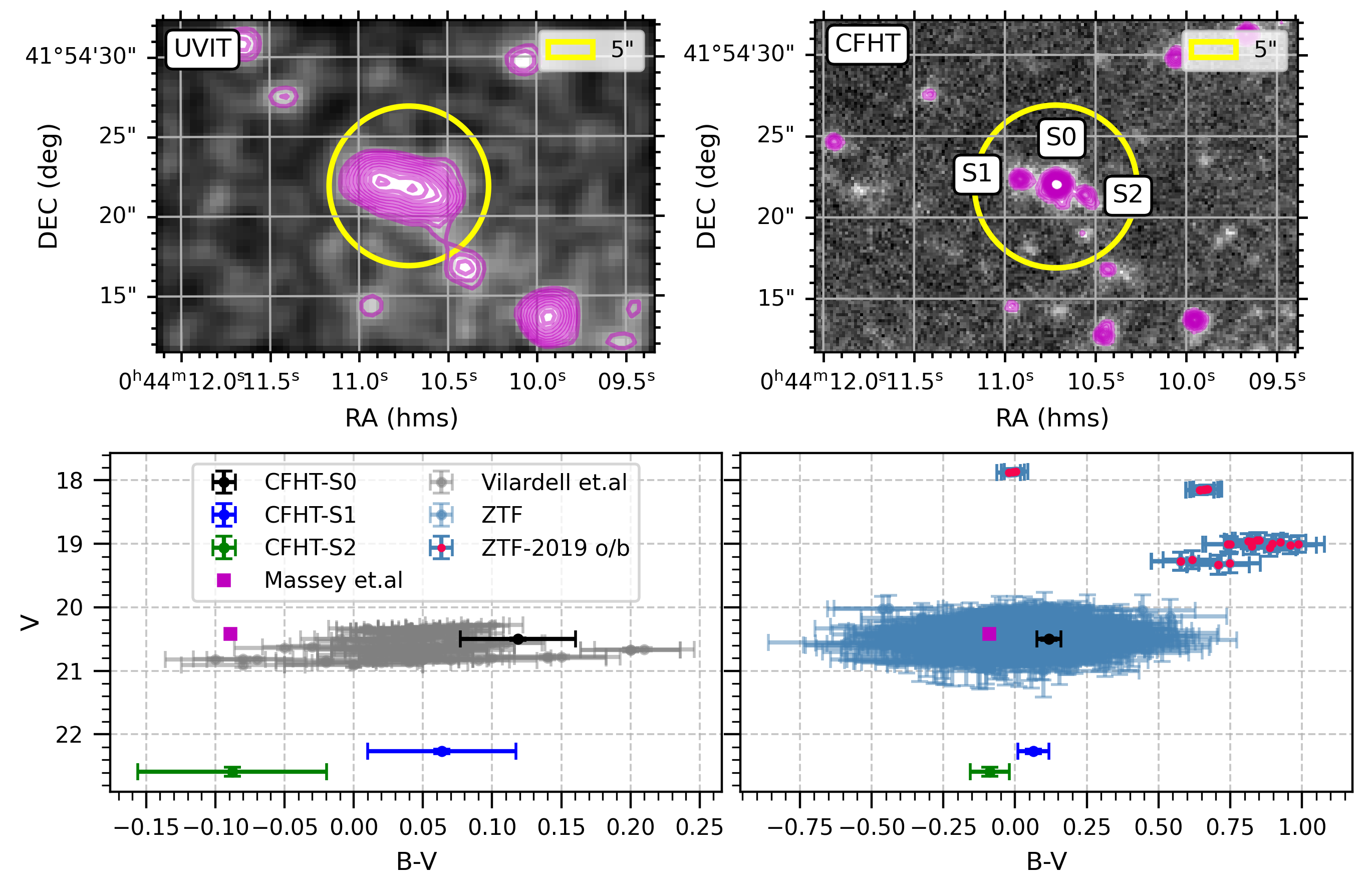}
    \caption{(Top-left): UVIT F148W image taken on 11th November 2022. (Top-right): Median combined CFHT u-band image. The yellow circle denotes a $5\arcsec$ radius centered on the reported nova position. Three sources within this region are labeled S1, S0, and S2 from left to right. (Bottom): Color–magnitude diagram of the three sources. Archival photometry from \citet{Vilardell}, \citet{Massey}, and ZTF corresponds to the position of CFHT-S0. Data points marked in red indicate the 2019 outburst.}
    \label{all_color}
\end{figure*}

\begin{figure*}
    \centering
    \includegraphics[width=1.0\textwidth]{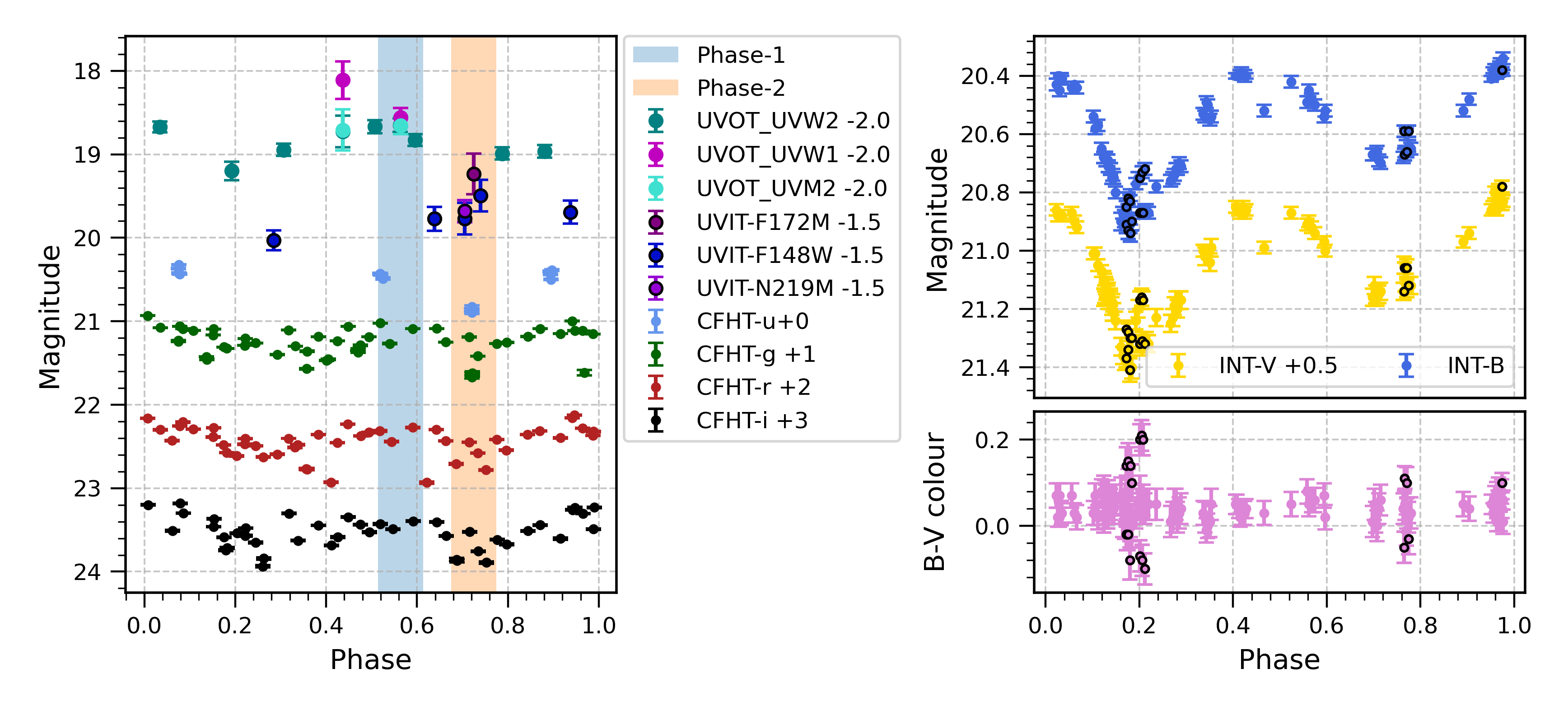}
    \caption{Phase-folded light curve. (Left): A period of 14.27 days is derived from the CFHT-$r$ data and the remaining data set is phase-folded to this period. Details of the CFHT, UVOT and UVIT data is available in Tables \ref{cfht-table},\ref{uvot-table},\ref{uvit-table} respectively. The two phase intervals highlighted were selected based on UV to optical coverage; these were subsequently used in the SED analysis. (Right): Archival light curve from \citet{Vilardell}, shown alongside the phase-dependent color variation. The black circles mark the points where B-V deviates $2\sigma$ from its median value.}
    \label{phase_lc}
\end{figure*}
\begin{figure*}[ht!]
    \centering
    \includegraphics[width=1.0\linewidth]{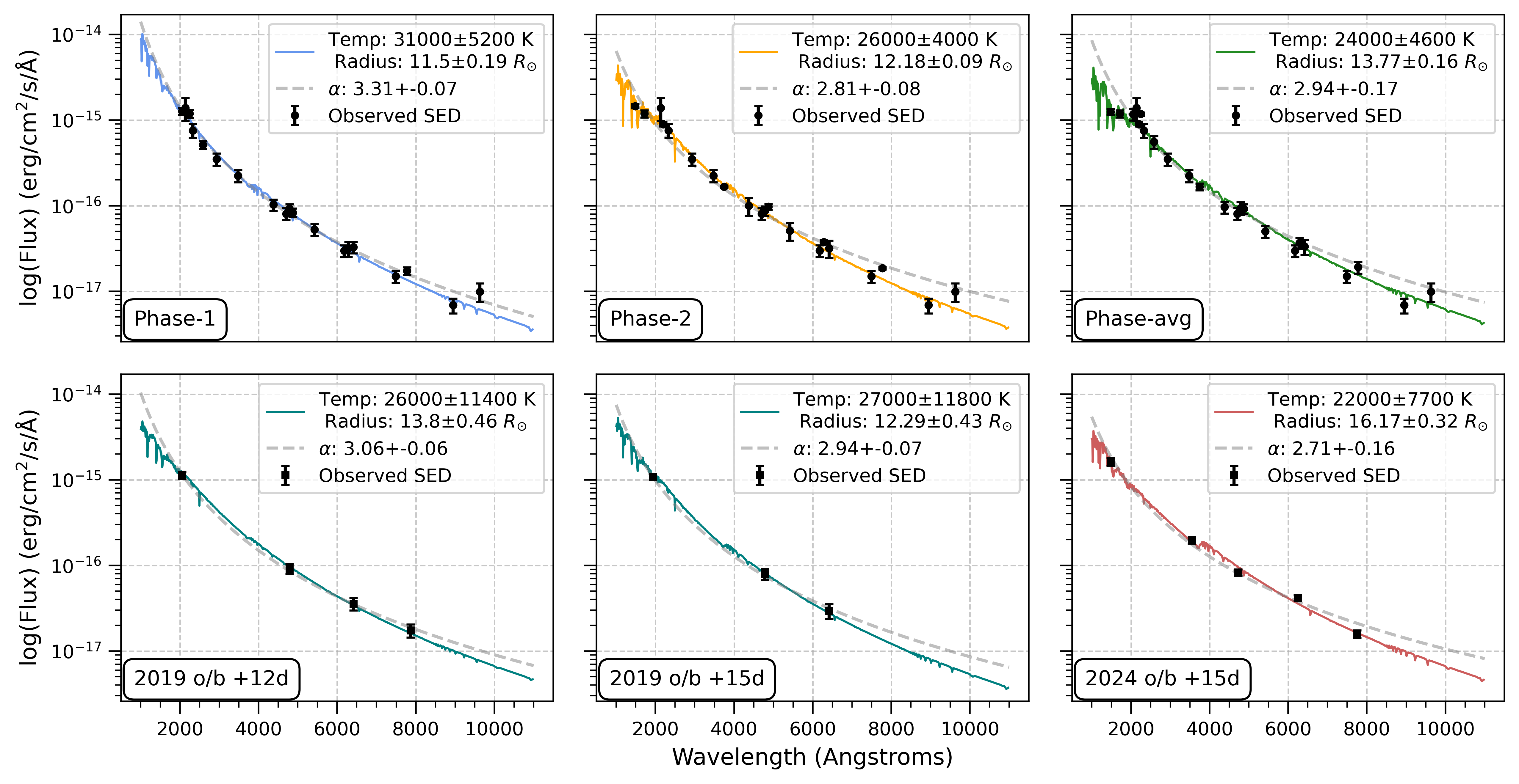}
    \caption{Observed SEDs (black marker) over-plotted with the best fit Kurucz model spectra (solid plot) and a power law fit (grey dash) with index alpha at different phases (see Fig. \ref{phase_lc}) and epochs (o/b = outburst). The temperature and radius are calculated from the best-fit Kurucz model.}
    \label{all-sed}
\end{figure*}
    
\section{Photometric Analysis During Outburst and Quiescence} \label{analysis}
We performed a comprehensive photometric analysis of M31N 2017-01e, capturing its behavior during both the 2024 and 2019 outbursts as well as its quiescent state. In the following sections, we describe the LCs during the outbursts, the period analysis and colors of the quiescent counterpart, and the color evolution during the 2019 outburst. We further present an analysis of the spectral energy distribution (SED) of the source at quiescence and post-outburst epochs, based on fits with Kurucz model spectra \citep{Kurucz}.

\subsection{Light Curve at Outburst}
We observed the 2024 outburst of the nova with the GIT, HCT, and UVIT and augmented with publicly available data from ZTF (see Fig. \ref{fig:lc} and Table-\ref{2024-opt}). The nova was observed at a peak magnitude of $m_{g^{\prime}} = 18.15 \pm 0.04$ and $m_{r^\prime} = 18.03 \pm 0.05$ with GIT on UT 6.85 August 2024, giving the closest constraint on the outburst time \citep{git-2024}. We calculate a $t_{2}(r^{\prime})=4.94\pm 0.20$ days and $t_{2}(g^{\prime})= 5.14\pm 0.24$ days, placing it in the \textit{very fast} speed class. We coordinated observations with the GIT, HCT and AstroSat $\sim\,15$ days post maximum. The nova was observed with AstroSat/SXT as the primary instrument in anticipation of detecting it in its SSS phase. But, no source was detected above the $2\sigma$ level, possibly due to low exposure. However, we do detect a source in the AstroSat/UVIT F148W band. These near-simultaneous observations in the optical and UV allow us to plot an SED at this epoch (see \S\ref{sed}).

Archival ZTF data covered this object in multiple epochs, including the 2019 outburst. Swift also followed up the 2019 outburst with both UVOT and XRT. We note that the ZTF-optical LC of this outburst is similar to that of the 2024 outburst. The UVOT photometry shows a similar trend as the optical LC. No X-ray source was detected in the XRT data above a $2.8\sigma$ sigma level, however the lack of a definitive soft X-ray signature is atypical post-nova outburst. There are two epochs of simultaneous UV and optical observations, at approximately 12 and 15 days after maximum, which we have used for SED analysis (see \S\ref{sed}).

We searched the CFHT archives between 2004-2012 (see Table-\ref{cfht-table}) for past outbursts but found none. However, we observe a variation in magnitude in the g,r,i bands in both CFHT and ZTF data which is analyzed in the next section.

\subsection{The Quiescent Counterpart}
We analyzed archival ZTF data to verify the periodicity reported for this source \citep{shafter-2022, Vilardell}. Data points corresponding to the 2019 outburst points were excluded, and a Lomb-Scargle periodogram was computed using the \texttt{LombScargle} class in \texttt{astropy}. To evaluate the significance of the observed peaks, we applied a $1\%$ false alarm probability (FAP) threshold, which quantifies the likelihood that a peak of similar power could arise purely from noise. We also use bootstrapping to get an estimate of uncertainty on the period. A dominant period of $14.30\pm3.25$ days and $14.29\pm6.6$ days was identified in ZTF-g and ZTF-r filters respectively. All data available for the source was phase-folded to the ZTF-g period, and a constant phase shift was applied to align them with the phased light curve reported by \citet{Vilardell}, to aid in comparing the modulation (see Fig.~\ref{lsp}). The CFHT data also shows a peak at a period of 14.27 days but below the $1\%$ FAP.

We searched the UVIT archives for the quiescent counterpart of the nova in UV. We detected an extended source at the location of the nova in both NUV and FUV filters (see Fig.~\ref{all_color}). The source is located $10.7^\prime$ from the center close to the edge, where distortion effects are more pronounced. However, neighboring sources in this region do not exhibit comparable elongation. Contour analysis of the nova region reveals that the elongated source can be resolved into two distinct sources. Despite a detection, this source was excluded from the UV catalog of novae in M31 using archival UVIT data \citep{uvit-m31} as sources overlapping within a $5\arcsec$ region were rejected.

We examined the high-resolution images in the CFHT/MegaPrime archives to investigate the origin of the elongated appearance of the source. The $u$-band epoch-wise and co-added images resolve three sources (S0, S1, S2) within this region (Fig.~\ref{all_color}). Although only the central source (S0) is clearly detected in individual $g$, $r$, and $i$ frames, the co-added images confirm all three across the filters. We performed grouped PSF photometry on UVIT and CFHT images, using \texttt{IRAF/nstar}, informed by source positions from CFHT. We verified flux residuals with \texttt{substar}, which shows nominal background variation, confirming reliable flux extraction. The sources S0, S1 and S1 are at a mean magnitude of $21.17 \pm 0.17$, $21.60 \pm 0.14$ and $22.44 \pm 0.23$ in the UVIT F147W filter. The full photometry for S0 is presented in Table \ref{uvit-table}.

S0 coincides precisely with the nova location and is the brightest, while S1 and S2 lie within $2\arcsec$ and appear faint, showing no discernible variability. Using the mean of the reported coordinates of M31N 2017-01e during its past outbursts, we calculate a separation of $0.16\arcsec$ from S0 in the CFHT-$u$ co-added image. This image has a limiting magnitude of 25, and contains 6445 sources within a $3\arcmin$ radius of S0, yielding a source density of $0.06~arcsec^{-2}$. Accordingly, the probability of detecting another source brighter than $u\sim25$ within the $0.16\arcsec$ uncertainty region is $0.53\%$. Repeating this calculation for the co-added images in the $g$, $r$, and $i$ filters gives probability of chance coincidence as $0.65\%$, $0.52\%$ and $0.30\%$ respectively. We next perform a color analysis to distinguish the nova system from field stars.

\subsection{Color Analysis} \label{colour}
We performed photometry on the three resolved sources within $5\arcsec$ of the nova location in the co-added CFHT images in the  $u$, $g$, $r$, and $i$ filters and calculated the corresponding colors. We note that S1 and S2 are distinctly fainter than S0 and appear to be unrelated to the nova.

We analyzed archival data from ZTF and other surveys assuming that the flux is dominated by S0. \cite{Vilardell} reported phased LCs and \cite{Massey} reported UBVRI photometry. We use these datasets to plot a $V \mathrm{~vs~} (B-V)$ color-magnitude diagram (CMD; see Fig \ref{all_color}). To enable direct comparison with CFHT and ZTF data, we employed color transformation equations by \cite{Jester} to derive $V$ magnitudes and $B-V$ colors from $g$ and $r$ band data. 
The $B-V$ colors reported in the literature include an average of the \citet{Vilardell} data set, $B-V=0.042$ \citep{shafter-2022}, though the data range from $-0.1$ to 0.2. The ZTF data set shows a broader spread at quiescence, ranging from $-0.5$ to 0.5, with an average of $B-V=0.003$. This larger spread likely arises from increased photometric uncertainties near the survey’s limiting magnitude. \citet{Massey} report $B-V=-0.089$, and we calculate $B-V=0.12\pm0.04$ from the co-added CFHT data. These differences are likely due to phase dependent variability and \slash or flux contamination from S1 and S2. However, all values fall within the broader range of the more densely sampled datasets. The ZTF dataset also includes photometry from the 2019 outburst, which is significantly brighter and redder than quiescent measurements, with  $B-V_{max} = 0.99\pm0.09$, consistent with the expected enhancement in H$\alpha$ emission during nova eruptions. We also note that this color change lasts for less than eight days, indicating the duration of the outburst. 

The $U-B$ value for S0 is found to be $-0.86$. Using this and the corresponding $B-V$ value, the reddening free Q parameter defined as $Q=(U-B) - 0.72(B-V)$ \citep{2004Ap.....47..499H,Orio} yields a value of $-0.94$. Q-values close to $-1$ are linked to spectral types earlier than B5, suggesting that the source is associated with a young stellar population. 

\subsection{SED Analysis}  \label{sed}
We phase folded all available data at quiescent epochs and constructed phased LCs for each filter (Fig. \ref{phase_lc}). To analyze the SED of the source during quiescence, we selected two orbital phases marked phase-1 and phase-2 in Fig. \ref{phase_lc}, that had data in both UV and optical. For each selected phase, photometric data within a phase interval of $\pm0.05$ were used, and a weighted average of the magnitudes for each filter within each bin was computed. The same procedure was applied across the entire orbital cycle to construct a phase-averaged SED for comparison.
In addition to the phased LC data, we incorporated catalog magnitudes in all three SEDs, with an additional uncertainty of 0.17 mag to account for the variability.  We considered the extinction maps by \cite{Montalto} and \cite{Draine}, which yield $\mathrm{A_V}$ of 0.82 and 1.69 (an additional factor of 0.5 was included in the dust mass to $\mathrm{A_V}$ conversion relation as mentioned in \citealt{uvit-m31}) respectively. We find that the fluxes obtained using $\mathrm{A_V} = 0.82$ are consistent with the spectrum of the source reported by \citet{shafter-2024}, and therefore adopt this value for all subsequent analysis. AB magnitudes were de-reddened using the extinction law of \citet{Cardelli} via the \texttt{extinction} Python package and converted to flux density ($\mathrm{F_{\lambda}}$). UVIT magnitudes were converted following the calibration of \citet{UVIT-cal}. 

We overplot the observed SEDs at different phases and epochs (see Fig. \ref{all-sed}) with the best least-squares fit Kurucz model spectra \citep{Kurucz}, and calculate stellar radius from the scale factor. We use bootstrapping to get an estimate of uncertainty in the best-fit model temperatures. Additionally, we fit a power law defined by $F_{\lambda}\propto \lambda^{-\alpha}$, where $\alpha$ is the spectral index.

At quiescence, the best fit Kurucz models span temperatures of $31000-26000$ K, radii $\sim$ 12-14 R$_{\odot}$ and $\alpha \sim 3$. The slightly higher temperature obtained at phase-1, is due to the absence of UVIT-F148W data, which otherwise constrains the fit to a lower temperature. The temperature estimates are consistent with that of a B-type star within the calculated uncertainties. However, the inferred radius or the extent of the emitting region is larger \citep{2013ApJS..208....9P}. While quiescent emission in novae systems is typically dominated by accretion discs, with $\alpha\sim2.33$ \citep{2002apa..book.....F,2013A&A...560A..49S,uvit-m31}, the power-law fits for this system deviate from this value. Furthermore, as also pointed out by \citet{shafter-2024}, an accretion disc alone cannot account for the observed quiescent luminosities. The observed SED aligns more closely with that of a B-type star.

The SEDs near outburst are consistent with those at quiescence. The  color evolution of the 2019 outburst indicates that the system returns to its quiescent colors within $\lesssim 8$ days, suggesting the emission at these epochs is dominated by a B-type stellar component.  

Additionally, we observe significant changes in the $B-V$ color around phases 0.2 and 0.8, which correspond to the LC minima (see Fig. \ref{phase_lc}). Near these phases both red-ward and blue-ward shift in color is observed.

\section{Discussion} \label{diss}
From deep CFHT imaging, we resolve three sources within $2\arcsec$ of the reported location of nova M31N 2017-01e (see Fig. \ref{all_color}). The central source, S0, is spatially coincident with the nova within sub-arcsecond accuracy and exhibits the periodic modulation reported by \citet{Vilardell}. The flanking sources, S1 and S2, are significantly fainter and bluer and show no variability in the CFHT $u$-band data, rendering them unlikely to be the progenitor.

\citet{shafter-2022} rule out the possibility of S0 being a galactic source. They reason that, since the system likely hosts a massive white dwarf accreting at high rates, emitting a quiescent luminosity of at least $\sim100~\mathrm{L_{\odot}}$, places S0 (with $\mathrm{m_V=20.4}$) at a distance of $\mathrm{100\,kpc}$, well outside the Milky Way. They also show that if the source were in M31, the standard nova model cannot account for its observed quiescent luminosity.

The photometric properties of S0 are strikingly atypical for nova secondaries. Known nova companions, late-type main sequence stars, sub-giants, or giants, exhibit typical $\mathrm{B-V}$ colors of $\sim0.36$, $0.64$, and $1.5$, respectively \citep{2012ApJ...746...61D}. In contrast, S0 shows $\mathrm{B-V = 0.12}$ and a reddening-free $\mathrm{Q = -0.94}$, indicating a hot, early-type star. SED fitting supports this, pointing toward a B-type stellar companion. Further, the modest outburst amplitude of $\sim 3$ mag deviates from the 6 mag minimum rise characteristic of recurrent novae \citep{2014ApJ...788..164P}. A B-type companion implies a mass ratio $q \gg 1$ with respect to the WD, which generally leads to unstable mass transfer via RLOF \citep{1967AcA....17..355P,2017MNRAS.471.4256V,2023A&A...669A..45T}. Wind accretion is unlikely to sustain the high accretion rates necessary for frequent eruptions \citep{2014A&A...564A..70K}. 

However, there exists a class of rapidly rotating early-type stars, known as Be stars, which episodically hosts a decretion disc, characterized by Balmer emission lines and infrared excess (see \citealt{2013A&ARv..21...69R} for review). These stars can form via binary interaction, wherein both binary components start off as early-type stars. Mass transfer from the primary, more massive component, via Roche Lobe overflow spins up the secondary, resulting in a rapidly rotating Be star and extending its lifetime on the main sequence. The primary subsequently evolves into either a Helium star, a Neutron Star (NS) or a WD (see \citealt{Raguzova} and references therein).
Such Be-compact object binaries are well established among Be/X-ray binaries (BeXRBs, see \citealt{Reig} for review), predominantly observed with NS companions in the Milky Way and Magellanic Clouds \citep{2015MNRAS.452..969C,2023A&A...677A.134N}. A smaller subset of Be star and WD systems (BeWD) have been identified through soft X-ray emission and early-type optical counterparts \citep[e.g.,][]{Kahabka-bewd1,Orio,Oliveira-bewd6,Sturm-bewd2,li-bewd3,mori-bewd3,Coe-bewd5,Kennea-bewd4,cxou-bewd7,ep-bewd8}. 

Observationally it has been noted that Be star discs can extend between $\sim10-100$ times the stellar radius ($R_\ast$) \citep{2013A&ARv..21...69R,2017A&A...601A..74K}, whereas, theoretical models show that BeWD systems are close binaries with components separated by less than $\mathrm{100\,R_\odot}$ \citep{Raguzova,zhue-bewd}. In the absence of a supernova kick in the WD formation, the WD orbit is coplanar and likely embedded in its companion's circumstellar disc (CSD) \citep{Raguzova}. Population synthesis studies confirm that both CO and ONe WDs in such systems can accrete at rates conducive to TNRs \citep{zhue-bewd}. At the same time, the dense CSD can obscure soft X-ray and UV emission \citep{Apparao}. However, their detection as X-ray transients implies the existence of some mechanism capable of dispersing the surrounding material or enabling accretion when the WD is not embedded in the disc.

BeXRBs have been observed to show X-ray flares when disc instabilities allow disc material to overflow the Roche lobe and accrete onto the compact object \citep{Okazaki-BeWD-ob}. \citet{Kennea-bewd4} point out that although this scenario of mass transfer is proposed for NS, it should hold good for any compact companion. \citet{Coe-bewd5} report similar behavior in BeWDs, with periodic flares and color changes suggesting disc expansion and episodic interaction with the WD.

The blue color and B-type stellar spectra being the closest match to the observed SED suggest that the optical companion of M31N 2017-01e is a Be-star. The radius inferred from the SEDs being larger than that of B-type stars is consistent with this scenario. All Be stars are found to be rapid rotators resulting in an extended emitting region around the equator in the form of a CSD, that can significantly contribute to the observed emission, particularly at longer wavelengths. The presence of this disc can be confirmed through deep infrared observation, which would be observed as an excess in flux compared to the expected B-type stellar emission (see Figure 3 in \citealt{cxou-bewd7}). In the absence of infrared detection for this object, the only available quiescent epoch Pan-STARRS-$y$ band ($\lambda_{\rm ref} = 9627.79\,\textup{\AA}$) point was included to check for this excess. We note a marginal rise at this wavelength from the best-fit model SED (see Fig.~\ref{all-sed}), although the deviation is not significant. These discs are known to change in size and can even dissipate completely over long timescales. An increase in disc size would manifest as an enhancement in redder flux. However, such changes are unlikely to occur over the timescale of a single orbital period. Interestingly, we do observe a variation in the B–V color for this object (see Fig.~\ref{all_color}), which may hint at disc-related changes. Furthermore, the spectrum of this source reported by \citet{shafter-2024}, consistent with a B-type star with H-alpha in emission along with a He I absorption feature, aligns with the Be-star scenario and adds further weight to this claim \citep{2021MNRAS.500.3926B}.

It is clear that only a massive WD ($M_{WD}\geq1.3 \mathrm{M_\odot}$) can show the recurrent nova outbursts observed in this system \citep{2018ApJ...860..110S}. Typical Be star masses range from $\mathrm{2\,to\,20\,M_{\odot}}$ \citep{2013A&ARv..21...69R}. Assuming the 14.3-day photometric period reflects the orbital period and $M_{\rm WD}$ = 1.3 M$_\odot$, for $M_{Be}$ in the range of 2-20 M$_\odot$, we derive a binary separation of $\sim$ 37-68 R$_\odot$ using Kepler's third law, well within the expected disc extent (i.e. $>10R_{\ast}$, here $>120\,\mathrm{R}_{\odot}$). We note that the assumed period of 14.3 days is on the shorter end for what is expected for such systems \citep{Raguzova,zhue-bewd}. In such case the CSD would appear as a pseudo Roche-lobe overflowing into the accretion disc of the WD, accreting sufficient mass for recurrent TNRs \citep{2013ApJ...777..136W,2014ApJ...793..136K,zhue-bewd}.

The luminosity of the Be star likely dominates the observed quiescence luminosity, which explains the observed low-amplitude outburst. The rapid evolution of the outburst light curve can be attributed to the small ignition mass required by a massive WD to trigger a TNR as in the case of the RRN M31N 2008-12a \citep{basu_21a}. There are two BeWD systems, namely  MAXI J0158–744 \citep{li-bewd3,mori-bewd3} and CXOU J005245.0–722844 \citep{cxou-bewd7,ep-bewd8}, that have been identified by a luminous soft X-ray burst accompanied by an enhancement in the optical bands by $\sim 0.5$ magnitude. These events are interpreted as TNR-induced nova explosions, followed by an SSS phase. Although the start of the optical outburst is poorly constrained for both cases, their optical light curves seem to decline rapidly, and the SSS phase is believed to start much sooner than that in typical novae systems.

M31N 2017-01e appears to lie between the BeWD and RRN classes of accreting WD systems. The low-amplitude outburst and its association with a B-type star are characteristic of BeWDs. In contrast, the light curve morphology, outburst spectrum, decline rate (i.e., the $t_2$ time), and ultra-short recurrence interval resemble those of the RRN. However, observations of the M31N 2017-01e outbursts to date have not detected a SSS phase, generally seen in other RRN. This absence may be due to unfavorable timing that missed the SSS phase, insufficient observational depth, or absorption by a dense circumstellar material. High-cadence, multi-wavelength monitoring during future outbursts will be essential to advance our understanding of such systems.

Given the extremely low probability of chance alignment with another source, S0 is likely associated with M31N 2017-01e in some manner. Our analysis assumes that S0 is the donor in the binary producing the recurrent outbursts, and that the observed periodicity reflects the orbital period of the system. Under these assumptions, the colors, SED, modest outburst amplitude, and binary configuration strongly support a BeWD scenario for M31N 2017-01e. However alternate scenarios such as, a configuration involving a triple system, should also be probed. In such a case S0 might not necessarily be the donor.  \citet{shafter-2022} also point out that the observed periodicity might not be orbital and actually arise from some other timescale within the system, such as accretion disc resonances, with the true orbital period potentially being longer. Understanding the actual system configuration would require high spatial and temporal resolution photometry and spectroscopy.

\section{Conclusion}
We report a multi-wavelength analysis in the UV and optical for M31N 2017-01e, the nova with the second shortest known recurrence period located in M31. The nova exhibits rapid evolution ($t_2 < 5$ days) and a low outburst amplitude ($\sim 3$ mag), distinguishing it from typical novae. Our analysis suggests the secondary is an early-type main sequence star, reminiscent of BeXRBs observed in the Milky Way and the Magellanic Clouds, quite unlike and unusual for a nova system. We propose that M31N 2017-01e hosts a massive WD in a binary with a Be star. This configuration naturally accounts for both the diminished outburst amplitude and short recurrence timescale. Infrared observations would be required to reveal the circumstellar disc and confirm the Be star nature. A systematic search for small amplitude variations in the optical bands can identify more such systems. Our findings highlight the potential diversity of nova progenitors and underscore the importance of multi-wavelength campaigns in unraveling their complex astrophysics. 

\section*{Acknowledgment}
We thank the anonymous reviewer for their insightful comments. 

The GROWTH India Telescope (GIT) is a 70-cm telescope with a 0.7-degree field of view, set up by the Indian Institute of Astrophysics (IIA) and the Indian Institute of Technology Bombay (IITB) with funding from  Indo-US Science and Technology Forum and the Science and Engineering Research Board, Department of Science and Technology, Government of India. It is located at the Indian Astronomical Observatory (Hanle), operated by IIA. We acknowledge funding by the IITB alumni batch of 1994, which partially supports operations of the telescope. Telescope technical details are available at \url{https://sites.google.com/view/growthindia/}. 

We thank the observing team of HCT. 

This publication uses the data from the AstroSat mission of the Indian Space Research Organisation (ISRO), archived at the Indian Space Science Data Centre (ISSDC). This work uses AstroSat/UVIT archival and ToO data. We thank the TAC for allotting us time to observe this object.

Based on observations obtained with MegaPrime/MegaCam, a joint project of CFHT and CEA/DAPNIA, at the Canada-France-Hawaii Telescope (CFHT) which is operated by the National Research Council (NRC) of Canada, the Institut National des Science de l'Univers of the Centre National de la Recherche Scientifique (CNRS) of France, and the University of Hawaii. The observations at the Canada-France-Hawaii Telescope were performed with care and respect from the summit of Maunakea which is a significant cultural and historic site. 

We acknowledge the use of public data from the Swift data archive. 
This research has made use of data and/or software provided by the High Energy Astrophysics Science Archive Research Center (HEASARC), which is a service of the Astrophysics Science Division at NASA/GSFC.

G.C.A. thanks the Indian National Science Academy (INSA) for support under the INSA Senior Scientist Programme.

\facilities{\textit{AstroSat} UVIT, \textit{Swift} UVOT and XRT, \textit{GIT}, \textit{HCT}, \textit{CFHT}}

\software{\texttt{CCDLAB} \citep{ccdlab}, 
         \texttt{IRAF v2.16} \citep{1986SPIE..627..733T},
         \texttt{HEASoft v6.34} \citep{2014ascl.soft08004N}, 
         \texttt{Python v3.10.12}, 
         \texttt{astropy v6.1.7} \citep{astropy:2022}
         \texttt{scipy v1.15.1} \citep{2020SciPy-NMeth}
         \texttt{NumPy v1.26.4} \citep{harris2020array},
         \texttt{pandas v2.2.0} \citep{pandas},
         \texttt{matplotlib v3.9.3} \citep{Hunter:2007},
         }
\nocite{*}
\bibliography{reference}

\appendix

\section{Tables}
\restartappendixnumbering
\begin{table*}[h]
    \centering
    \caption{AstroSat/UltraViolet Imaging Telescope archival and 2024 outburst data log and magnitudes}
    \begin{tabular}{ccccc}
    \toprule
    Observation ID & Observation UT & Filter & Exposure Time(s) & Magnitude \\
    \midrule
    A04\_022T04\_9000001746 & 2017-12-03.70 & F148W & 3255 & $ 21.27 \pm 0.19 $ \\
    A04\_022T04\_9000001746 & 2017-12-03.70 & N219M & 14466 & $ 21.18 \pm 0.13 $ \\
    A04\_022T04\_9000001746 & 2017-12-03.98 & F172M & 9656 & $ 20.74 \pm 0.24 $ \\
    T03\_262T01\_9000003988 & 2020-10-30.28 & F148W & 6631 & $ 21.19 \pm 0.14 $ \\
    T03\_262T01\_9000003988 & 2020-11-10.76 & F148W & 8170 & $ 21.00 \pm 0.19 $ \\
    T04\_072T01\_9000004780 & 2021-11-23.22 & F148W & 25505 & $ 22.27 \pm 0.10 $ \\
    T05\_058T01\_9000005414 & 2022-12-07.24 & F148W & 7872 & $ 21.27 \pm 0.14 $ \\
    A13\_003T01\_9000006002 & 2023-12-23.31 & F148W & 11342 & $ 21.53 \pm 0.12 $ \\
    T05\_221T01\_9000006404 & 2024-08-22.15 & F148W & 642 & $ 21.00 \pm 0.27 $ \\
    \bottomrule
    \end{tabular}
    \label{uvit-table}
\end{table*}

\begin{table}[h]
    \centering
    \caption{GROWTH-India Telescope and Himalayan Chandra Telescope photometry for 2024 outburst of M31N 2017-01e}
    \begin{tabular}{cccc}
    \toprule
    Observation UT & Telescope/Filter & Exposure Time(s) & Magnitude \\
    \midrule
    2024-08-06.84 & GIT-$g^\prime$ & 360 & $ 18.10 \pm 0.03 $ \\
    2024-08-06.85 & GIT-$r^\prime$ & 360 & $ 17.88 \pm 0.04 $ \\
    2024-08-09.73 & GIT-$i^\prime$ & 400 & $ 19.53 \pm 0.18 $ \\
    2024-08-09.74 & GIT-$r^\prime$ & 400 & $ 19.35 \pm 0.08 $ \\
    2024-08-09.74 & GIT-$g^\prime$ & 400 & $ 19.58 \pm 0.07 $ \\
    2024-08-10.77 & GIT-$g^\prime$ & 900 & $ 19.74 \pm 0.05 $ \\
    2024-08-10.80 & GIT-$i^\prime$ & 900 & $ 19.82 \pm 0.08 $ \\
    2024-08-10.80 & GIT-$r^\prime$ & 900 & $ 19.50 \pm 0.05 $ \\
    2024-08-13.77 & GIT-$g^\prime$ & 600 & $ 20.18 \pm 0.11 $ \\
    2024-08-13.74 & GIT-$i^\prime$ & 900 & $ 20.19 \pm 0.13 $ \\
    2024-08-13.78 & GIT-$r^\prime$ & 900 & $ 20.45 \pm 0.10 $ \\
    2024-08-15.82 & GIT-$g^\prime$ & 900 & $ 20.36 \pm 0.06 $ \\
    2024-08-15.81 & GIT-$i^\prime$ & 900 & $ 20.56 \pm 0.11 $ \\
    2024-08-15.83 & GIT-$r^\prime$ & 900 & $ 20.57 \pm 0.07 $ \\
    2024-08-22.83 & GIT-$r^\prime$ & 1800 & $ 20.28 \pm 0.08 $ \\
    2024-08-22.91 & GIT-$i^\prime$ & 3240 & $ 20.67 \pm 0.11 $ \\
    2024-08-22.76 & HCT-$u^\prime$ & 1620 & $ 20.41 \pm 0.09 $ \\
    2024-08-22.80 & HCT-$g^\prime$ & 1020 & $ 20.42 \pm 0.09 $ \\
    \bottomrule
    \end{tabular}
    \label{2024-opt}
\end{table}

\begin{table}[h]
    \centering
    \caption{Archival Canada-France-Hawaii Telescope data log and magnitudes (Full table is available in the machine readable format) }
    \begin{tabular}{ccccc}
    \toprule
    Observation ID & Observation UT & Filter & Exposure Time(s) &  Magnitude \\ 
    \midrule
    04BH54 & 2004-08-07.60 & r.MP9601 & 400 & $ 20.125 \pm 0.007 $ \\
    04BH54 & 2004-08-07.61 & g.MP9401 & 530 & $ 20.114 \pm 0.005 $ \\
    04BH54 & 2004-08-07.62 & i.MP9701 & 530 & $ 20.232 \pm 0.008 $ \\
    04BH54 & 2004-08-09.58 & r.MP9601 & 400 & $ 20.209 \pm 0.006 $ \\
    04BH54 & 2004-08-09.58 & g.MP9401 & 530 & $ 20.090 \pm 0.004 $ \\
    04BH54 & 2004-08-09.59 & i.MP9701 & 530 & $ 20.300 \pm 0.006 $ \\
    04BH54 & 2004-08-10.53 & r.MP9601 & 400 & $ 20.386 \pm 0.007 $ \\
    04BH54 & 2004-08-10.54 & g.MP9401 & 530 & $ 20.166 \pm 0.004 $ \\
    04BH54 & 2004-08-10.55 & i.MP9701 & 530 & $ 20.462 \pm 0.008 $ \\
    04BH54 & 2004-08-11.54 & r.MP9601 & 400 & $ 20.406 \pm 0.006 $ \\
    04BH54 & 2004-08-11.54 & g.MP9401 & 530 & $ 20.207 \pm 0.004 $ \\
    \vdots & \vdots & \vdots & \vdots & \vdots  \\
    \vdots & \vdots & \vdots & \vdots & \vdots \\
    \vdots & \vdots & \vdots & \vdots & \vdots \\
    \bottomrule
    \end{tabular}
    \label{cfht-table}
\end{table}

\begin{table}[h]
    \centering
    \caption{Swift/UltraViolet Optical Telescope archival data log and magnitudes for the 2019 outburst of M31N 2017-01e}
    \begin{tabular}{ccccc}
    \toprule
    Observation ID & Observation UT & Filter & Exposure Time(s)& Magnitude \\
    \midrule
    00037715001 & 2008-07-07.68 & UVW2 & 289 & $ 20.73 \pm 0.19 $ \\
    00037715001 & 2008-07-07.68 & UVM2 & 289 & $ 20.71 \pm 0.25 $ \\
    00037715001 & 2008-07-07.69 & UVW1 & 285 & $ 20.11 \pm 0.22 $ \\
    00037715002 & 2008-08-21.39 & UVW2 & 1288 & $ 20.65 \pm 0.08 $ \\
    00037715002 & 2008-08-21.40 & UVM2 & 1288 & $ 20.65 \pm 0.11 $ \\
    00037715002 & 2008-08-21.40 & UVW1 & 1135 & $ 20.56 \pm 0.11 $ \\
    00033061001 & 2013-12-27.29 & UVW2 & 1714 & $ 21.20 \pm 0.11 $ \\
    00033061002 & 2013-12-31.79 & UVW2 & 1698 & $ 20.67 \pm 0.08 $ \\
    00033061003 & 2014-01-04.80 & UVW2 & 3243 & $ 20.99 \pm 0.07 $ \\
    00033061004 & 2014-01-08.33 & UVW2 & 2818 & $ 20.67 \pm 0.06 $ \\
    00033061005 & 2014-01-12.23 & UVW2 & 3069 & $ 20.95 \pm 0.07 $ \\
    00033061006 & 2014-01-16.37 & UVW2 & 2911 & $ 20.83 \pm 0.07 $ \\
    00033061007 & 2014-01-20.43 & UVW2 & 3168 & $ 20.96 \pm 0.07 $ \\
    00012018001 & 2019-09-27.71 & UVW2 & 1962 & $ 19.88 \pm 0.05 $ \\
    00012018002 & 2019-10-05.75 & UVW2 & 1443 & $ 20.87 \pm 0.10 $ \\
    00012018003 & 2019-10-08.28 & UVW2 & 1736 & $ 20.78 \pm 0.09 $ \\
    00012018005 & 2019-10-16.24 & UVW2 & 1291 & $ 20.55 \pm 0.09 $ \\
    00012018008 & 2019-10-24.53 & U & 2270 & $ 20.35 \pm 0.08 $ \\
    \bottomrule
    \end{tabular}
    
    \label{uvot-table}
\end{table}

\end{document}